\documentstyle[prl,aps,epsf]{revtex}
\draft
\begin{document}
\twocolumn[\hsize\textwidth\columnwidth\hsize\csname @twocolumnfalse\endcsname

\title{Depinning and Dynamics of Systems with Competing 
Interactions in Quenched Disorder} 
\author{C. Reichhardt, C.J. Olson, I. Martin, and A.R.~Bishop} 
\address{ 
Center for Nonlinear Studies, 
Theoretical Division and  Applied 
Physics Division, 
Los Alamos National Laboratory, Los Alamos, NM 87545}

\date{\today}
\maketitle
\begin{abstract}
We examine the depinning and driven dynamics of a system in
which there is a competition between long range Coulomb repulsive and 
short range attractive interactions. 
In the absence of disorder the system forms Wigner crystal,
stripe and clump phases as the attractive interaction is increased. 
With quenched disorder, these phases are fragmented and 
there is a finite depinning threshold. The stripe phase is the most
strongly pinned and shows hysteretic transport properties.  
At higher drives beyond depinning, a dynamical reordering 
transition occurs in all the phases, which is associated with a characteristic
transport signature.
\end{abstract}
\vspace{-0.1in}
\vspace{-0.3in}

\vskip2pc]
\narrowtext
Two-dimensional systems in which there is a competition between
long-range repulsion and short range attraction exhibit a 
remarkable variety of patterns such as stripes, bubbles,
and labyrinths \cite{Andelman}. 
Such  systems include magnetic films \cite{Seul}, 
Langmuir monolayers, polymers, gels, and water-oil mixtures \cite{Gelbart}.  
It has been proposed that similar competing interactions
can arise in two-dimensional electron systems leading to 
stripes, clumps \cite{Electron} and liquid crystalline electron 
states \cite{Kivelson}. Stripe and other charge-ordered phases 
in metal oxides are sometimes modeled as systems with competing long
range repulsion and short range attraction \cite{Bishop}.       
In many of these systems quenched disorder from the underlying
substrate may be present; however, it is not known how this disorder 
would affect the structure and dynamics of these systems.
Quenched disorder can strongly alter the transport properties, 
producing a pinning effect in
which a finite driving force must be applied before net motion occurs. 

The effects of quenched disorder on 
driven elastic media have been studied extensively in the context of
vortices in superconductors \cite{Vinokur,Giamarchi,Olson,Cha,Shobo,Pardo}, 
Wigner crystals \cite{Wigner} and charge density waves (CDWs) \cite{Fisher}.
A remarkable wealth of nonequilibrium behaviors arise,
one of the most intriguing of which is {\it dynamic reordering}. 
At low drives above depinning, the elastic system can break up
and flow plastically. For increasing drive, the effects 
of the quenched disorder are partially reduced, allowing the
elastic interactions to dominate.  The system then
orders or partially orders to a moving crystal or smectic. 
Dynamical reordering has been studied theoretically \cite{Vinokur,Giamarchi}, 
experimentally \cite{Shobo,Pardo} and in simulations \cite{Olson} 
in vortex matter, CDW systems \cite{Fisher} and driven Wigner crystals 
\cite{Wigner}. An open question is whether dynamic reordering is universal to 
other types of systems with quenched disorder, particularly those with 
competing interactions.      

The study of transport in systems with competing
interactions and quenched disorder is of considerable value
since in many  physical systems it is not possible to
probe the microscopic behavior directly.  Instead transport signatures
such as $I-V$ characteristics and conduction
noise are measured. 
Recently, highly nonlinear and hysteretic transport was observed
for two dimensional electrons in the reentrant integer quantum Hall
state, and it was suggested that this is a signature
of the depinning of some form of 
charge ordered state \cite{Lilly}. We note that simulations for driven 
vortex matter \cite{Olson} and Wigner crystal states \cite{Wigner} in 
2D have produced {\it non hysteretic} transport curves; however, it is not 
known if the depinning of other ordered states such as stripes or
clumps is hysteretic. 

In order to examine the transport characteristics as well as to determine
whether dynamic reordering occurs in
this class of system, we have conducted numerical simulations of systems 
of particles that have a competing long-range Coulomb repulsion and
a short range attraction. Our model is  similar to one introduced by 
Stojkovic {\it et al.} \cite{Bishop} to examine charge ordering in
metal oxides.  In our model, as a function of increasing attractive
interaction, we find three generic phases in the absence of 
quenched disorder similar
to those observed in \cite{Bishop}:
Wigner crystal, stripe, and clump.
The addition of disorder affects the stripe phase most strongly,
fragmenting the stripes and 
producing a large depinning threshold 
and nonlinear $I-V$ characteristics.
The depinning threshold drops sharply upon entering the clump phase.
The $I-V$ curves are 
hysteretic in the stripe region and in a portion of the clump region 
where the disorder fragments the clumps. The initial depinning of 
the fragmented stripe and clump phase is plastic, but at higher drives 
there is a dynamical reordering, where ordered stripes that
are aligned with the drive form and the clumps reform. Characteristic
signatures of
the reordering appear as features in the $I-V$ curves as well as the 
conduction noise, which shows a $1/f^{2}$ characteristic in the moving
fragmented stripe and clump phase, and a washboard 
periodic signal in the moving
reordered states. 

We model a 2D system of overdamped 
interacting particles that have a long-range 
Coulomb repulsion and a short range exponential attraction. 
The equation of motion for a particle $i$ is 
$ {\bf f}_{i} =  
{\bf f}_{ii} + {\bf f}_{p} + {\bf F}_{d} = 
\eta{\bf v}_{i}$, where
the damping term $\eta=1$.
The force from the other particles is
 ${\bf f}_{ii} = -\nabla U(r)$, where 
$$
U(r) = 1/r - B \exp(-\kappa r).
$$
$\kappa=0.25$ is the inverse range, and
the parameter B is used to vary 
the relative strength of the attractive interaction.
The repulsive Coulomb term, treated with a summation method \cite{Gronbech},
dominates at small and large $r$.
The force from the quenched disorder ${\bf f}_{p}$ comes from 
randomly placed attractive parabolic pins of
maximum strength $f_{p}$. 
The driving term ${\bf F}_{d}$ 
is increased from 
0 to $0.08$ in our units 
in increments of $0.0002$, and the sample is held at each drive 
increment for $8\times 10^4$ time steps to ensure a steady state.
We measure
the time-averaged particle velocity $<V_{x}>$ and its derivative $dV/dF$.
The system is initially prepared in a high temperature molten state
and annealed to $T = 0$. After annealing the driving force is applied.  
For the work presented here we keep
$\kappa$ and the particle density fixed and vary $B$. Changing
particle density and $\kappa$ do not affect the qualitative results. 

We first investigate the transport properties for different phases as
a function of $B$. For the parameters we present here, 
in the absence of disorder we find a Wigner crystal for $B = 0$.  As $B$ is 
increased the lattice becomes increasingly distorted until for 
$B = 0.25$ to $B < 0.325$ stripes form, and for $B \ge 0.325$ clumps form. In
Fig.~1(a) we show representative velocity vs driving force curves
for the system with quenched disorder for the Wigner crystal, 
stripe and clump phases. 
The curves for the Wigner and stripe phases are highly
nonlinear and S-shaped.
In addition the pinning is much more effective for the stripe phase than
the Wigner crystal as indicated by the larger depinning threshold.  
In Fig.~1(b) we present the corresponding
$dV/dF$ curves. 
The initial depinning for $B < 0.35$ is plastic which coincides with
the onset of strong peaks in the $dV/dF$ curves.  
The stripe phase shows a second small peak in the 
$dV/dF$ curve, in addition to the large plastic peak. 

In Fig.~1(c) we show the depinning threshold vs B, obtained from 
the velocity curves where we take depinning to occur at 
a velocity threshold of $0.002$. The figure also 
indicates where the Wigner, stripe, and clump phases occur as a function of
$B$ without disorder. 
The susceptibility of the stripe state 
to the quenched disorder and the corresponding increase in
the depinning threshold can be viewed as a consequence of the softening of the
system due to self-induced disorder. Recently Schmailian and Wolynes 
\cite{Wolynes}
proposed that for systems with competing interactions without quenched 
disorder, in particular stripe generating systems, self-generated
glass or disordering transitions occur due to frustration
\cite{Wolynes}.  A disordered system is
{\it soft}, allowing individual particles to be displaced without costing 
elastic energy, so that particles can find optimal pinning
sites. Conversely a system with strong elastic interactions such as a
crystal is very stiff, preventing particles from being randomly
displaced to the optimal pinning sites.  In our system, for low B 
the Coulomb interaction dominates and the system forms 
a distorted Wigner crystal
state. As $B$ increases there is increasing competition between the
repulsive and attractive interactions. 
In the stripe phase the elasticity is lowest,
and the system can most easily adjust to the pinning. 
This implies that well defined stripe structures will be easily 
destroyed in systems where quenched disorder is present. For increasing
$B$ the short-range attraction becomes dominant and clumps form. 
The pinning is strongly reduced in the clump region due to the increase in 
elasticity from three effects:  once the particles are in the clump
phase, individual particles cannot leave the clump to take advantage of
the pinning;  within the clumps the particles form crystal 
structures which create an intraclump elasticity;  additionally, since 
Coulomb interactions dominate at long range, the clumps themselves
form a partially ordered Wigner crystal with an additional elasticity. 
An increase in the pinning due to the softening of an elastic media  
has also been evoked for explaining the peak effect phenomena observed
in vortex matter in type II superconductors where an increase of the pinning
occurs as a function of field or temperature \cite{Shobo}.   
Simulations for vortex matter where the lattice is softened have shown 
that this softening will lead to a smooth increase in the depinning 
threshold as well as an onset and enhancement of the plastic flow 
region \cite{Cha}.  

In Fig.~2 we show a series of snapshots for different values of the
applied driving force for the system 
at $B=0.29$ in the stripe phase to indicate the correlation between
the various features in the velocity force curves and
the microstructure of the moving system. 
Fig.~2(a) shows the $F_{d} = 0$ disordered stripe phase. Here
only stripe fragments can be seen. Just above depinning, at $F_{d} = 0.021$ 
in Fig.~2(b), the system is in plastic flow and the stripe structures
are almost completely destroyed. For drives up to the first peak in the
$dV/dF$ curve in Fig.~1(b), the system is strongly disordered,
while above the peak the system begins to reorder, as seen in 
Fig.~2(c,d) for $F_{d} = 0.029$ and $0.04$ where stripe
segments reform. For $F_{d} = 0.045$ [Fig.~2(e)]
an {\it aligned stripe} phase forms. 
For drives beyond the second peak in $dV/dF$,
as shown in Fig.~2(f) ($F_{d} = 0.06$),  
the stripe structures are very well formed. 
The second peak in $dV/dF$ appears because the reordering transition
leads to a change in the effectiveness of the pinning, from stronger
pinning (and slower-moving particles)
in the disordered state to weaker pinning (and faster-moving particles at
the same drive) in the ordered stripe state.

In Fig.~3(a-f) we show the reordering dynamics 
for $B = 0.325$, where in the absence of quenched disorder the 
systems forms the
clump state. For this value of disorder the 
$F_{d} = 0.0$ clump state is fragmented
as seen in Fig.~3(a).
Above depinning, as in Fig.~3(b,c) ($F_{d} = 0.018, 0.02$), the  
clumps are further fragmented. In Fig.~3(d) and Fig.~3(e) for $F_{d} =
0.03$ and $0.035$ the clumps began to reorder. In Fig.~3(f) for $F_{d} =
0.06$ the clumps are completely reordered.  

For $B > 0.35$, the clumps are no longer fragmented by the 
quenched disorder at $F_{d}=0$, and the depinning transition is
elastic, with the entire clump structure depinning as a
unit, and with greatly diminished peaks in the $dV/df$ curves.
For $B < 0.25$ the system forms a disordered Wigner crystal 
which depins plastically and forms a moving smectic, as previously studied
in Ref~\cite{Wigner}. 

In Fig.~4(a) we show that only certain phases have hysteretic properties.
The velocity-force curve for the stripe phase shows hysteresis
which corresponds to the ordered stripe structure, as seen in Fig.~2(e),
``supercooling'' on the downward drive sweep. 
We do not observe any hysteresis for the 
elastically depinning clump phases at $B>0.35$,
since the moving state and the pinned state are structurally 
the same. Additionally we find no hysteresis for the 
Wigner crystal regions $ B < 0.25$, which is in agreement with
previous work on driven Wigner crystals
as well as previous work for vortex matter in 2D. 

We have measured the power spectra $S(f)$ of the velocity fluctuations
at fixed drive values, and find two general forms.
In the moving disordered or plastic flow
phases, we observe a broad band signal with a $1/f^{\alpha}$
characteristic.  In the reordered moving phases we observe narrow band 
peaks.  In Fig.~4(b) we show $S(f)$ for the
stripe phase in the plastic flow region and the reordered region. 
For the plastic flow regime a $1/f^{2}$ signal can be seen, while for
the reordered phase the noise power is considerably reduced and a series 
of narrow band peaks are observable. For the moving Wigner crystal we
observe only a single narrow band peak, produced by the washboard
signal of the moving lattice of frequency $\omega = v/a$, where $v$ is
the velocity of the lattice and $a$ is the lattice constant. This has
also been seen for vortices moving in an ordered phase
\cite{Olson}. For the reordered
stripe and clump phase, we observe a series of peaks, as shown. The
series of peaks rather than one washboard signal arise
due to the combination of the macro-structural ordering of the 
moving stripe and clump phases,
as well as the internal ordering of the particles
within the stripes and clumps. 
These additional internal
frequencies are at much higher frequencies than the washboard signal
since the internal structure lattice constant is much smaller than the 
macrostructure lattice constant. 

We briefly comment that we have also performed simulations for different 
values of disorder. We find that the peak in the pinning force always 
corresponds to the stripe phase. For larger disorder the peak broadens
gradually. For different values of $\kappa$ or for different 
density of particles we observe  that the onset of the different
phases occurs for different values of $B$ than presented here; however,  
the qualitative behaviors are unchanged. 

In conclusion we have used numerical simulations to examine the
depinning and dynamics of a system with competing long-range repulsion
and short range attraction where in the absence of quenched disorder
we find a Wigner crystal, stripe and clump phase. With quenched
disorder these phases become partially disordered.  The 
stripe phase is most strongly affected by the quenched disorder 
and has the highest depinning threshold due to the self-induced 
frustration. It may be difficult to observe
well defined stripe structures in
systems with quenched disorder. For increasing drives we find that
these phases can exhibit a dynamical reordering transition. For
the fragmented stripe phase the transition is hysteretic with 
the stripes becoming highly ordered and aligned with the drive.     
The enhanced pinning and hysteresis we observe resemble those found
recently in experiments on two-dimensional electron
systems \cite{Lilly}.
We also demonstrate that the reordering transitions can be inferred
from characteristics in the $I-V$ curves and noise spectra.  
The dynamics and the mesoscopic patterns we observe
should be generic to the general class of systems
with competing long-range and short-range interactions in quenched disorder.

Acknowledgments---We thank A. Castro-Neto for useful discussions. 
This work was supported by the US Department of Energy under
contract W-7405-ENG-36.

\begin{figure}
\center{
\epsfxsize=3.3in
\epsfbox{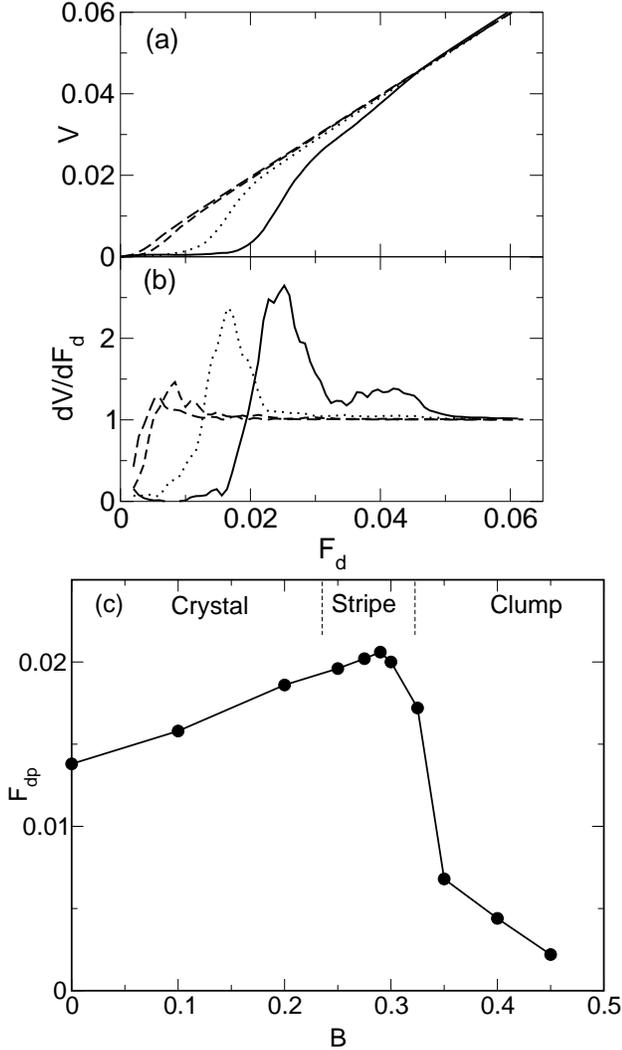}}
\caption{
(a) Velocity $V$ vs applied drive $F_{d}$ for 
$B = 0.29$ (solid line) stripe phase, $B=0.0$ (doted line) Wigner crystal,
$B=0.35$ (dashed) clump phase, and $B=0.4$ (long-dashed) clump phase. 
(b) The corresponding dV/dF curves. (c) Depinning force vs B. 
Labels indicate where the Wigner crystal, stripes, and clumps form 
in the absence of quenched disorder. 
}
\end{figure}

\begin{figure}
\center{
\epsfxsize=3.5in
\epsfbox{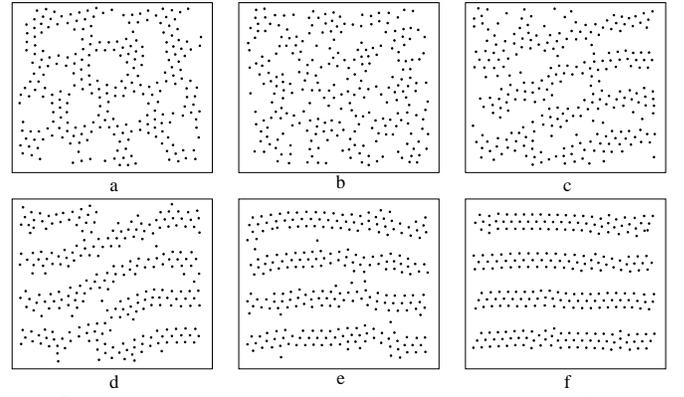}}
\caption{
Individual snapshots for increasing applied drive for 
$B = 0.29$. (a) $F_{d} = 0.0$, (b) $F_{d} = 0.021$, (c)
$F_{d} = 0.029$, (d) $F_{d} = 0.04$, (e) $F_{d} = 0.045$ and 
(f) $F_{d} = 0.06$.
}
\end{figure}

\begin{figure}
\center{
\epsfxsize=3.5in
\epsfbox{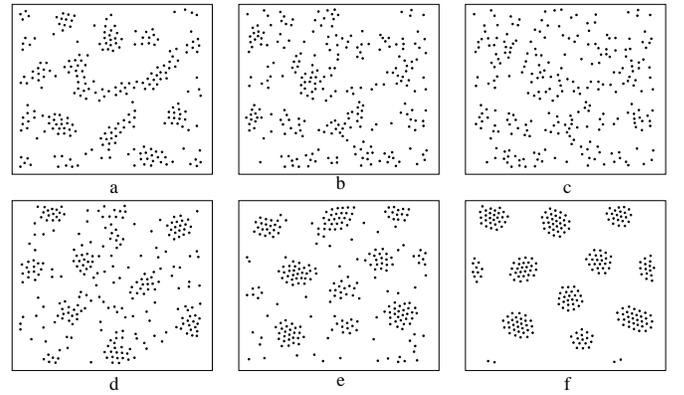}}
\caption{
Individual snapshots for increasing applied drive for 
$B = 0.325$ for (a) $F_{d} = 0.0$, (b) $F_{d} = 0.018$,
(c) $F_{d} = 0.02$, (d) $F_{d} = 0.03$, (e) $F_{d} = 0.035$, and 
(f) $F_{d} = 0.06$. 
}
\end{figure}

\begin{figure}
\center{
\epsfxsize=3.5in
\epsfbox{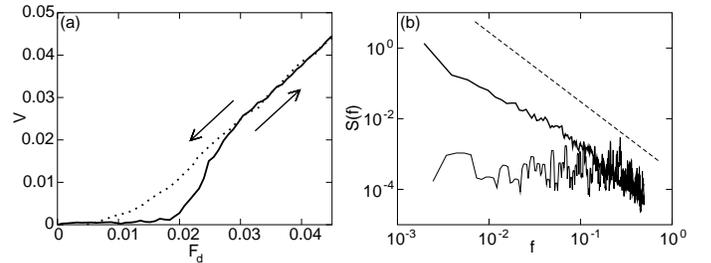}}
\caption{
(a) The hysteretic response in velocity force curves for the stripe
phase $B = 0.29$. 
(b) Noise curves for driven stripe phase 
$B = 0.29$ (thick line) in the plastic flow phase $F_{d} = 0.021$ showing
a broad band signature. The dashed line is the $1/f^{2}$ curve. 
Thin line is the noise curve for the reordered state $F_{d} = 0.06$.  
}
\end{figure}


\begin{references}

\bibitem{Andelman}  
For a review see M.~Seul and D.~Andelman, Science {\bf 267}, 476 (1995). 

\bibitem{Seul}
M.~Seul and R.~Wolfe, Phys.~Rev.~A, {\bf 46}, 7519 (1992).

\bibitem{Gelbart}
W.M.~Gelbart and A.~Ben Shaul, J.~Phys.~Chem. {\bf 100}, 13 169 (1996). 

\bibitem{Wolynes}
J.~Schmalian and P.G.~Wolynes, Phys.~Rev.~Lett.~{\bf 85}, 836 (2000). 

\bibitem{Electron}
A.A.~Koulakov, M.M.~Fogler, and B.I.~Shklovskii, Phys.~Rev.~Lett. {\bf 76},
499 (1996);
M.M.~Fogler, A.A.~Koulakov and B.I.~Shklovskii, Phys.~Rev.~B {\bf 54}, 1853
(1996).
 
\bibitem{Kivelson}
E.~Fradkin and S.A.~Kivelson, Phys.~Rev.~B, {\bf 59}, 8065 (1999).

\bibitem{Bishop}
B.P.~Stojkovic {\it et al.}, Phys.~Rev.~Lett.~{\bf 82}, 4679 (1999);
Phys.~Rev.~B {\bf 62}, 4353 (2000), and references therein.

\bibitem{Vinokur}
A.E.~Koshelev and V.M.~Vinokur, Phys.~Rev.~Lett.~{\bf 73}, 3580 (1994). 

\bibitem{Giamarchi}
P.~Le Doussal and T.~Giamarchi, Phys.~Rev.~B {\bf 57}, 11 356 (1998);
L.~Balents, M.C.~Marchetti and L.~Radzihovsky {\bf 57}, 7704 (1998);
S.~Scheidl and V.M.~Vinokur, Phys.~Rev.~E {\bf 57}, 2574 (1998).

\bibitem{Olson}
K.~Moon, R.T.~Scalettar, and G.T.~Zimanyi, Phys.~Rev.~Lett.~{\bf 77}, 2778
(1996); C.J.~Olson, C.~Reichhardt, and F.~Nori, Phys.~Rev.~Lett.~{\bf 81},
3757 (1998); A.B.~Kolton, D.~Dominguez, and N.~Gr{\o}nbech-Jensen,
Phys.~Rev.~Lett.~{\bf 83} 3061 (1999). 

\bibitem{Cha}
Min-Chul Cha and H.A.~Fertig, Phys.~Rev.~Let.~{\bf 80}, 3851 (1998);
C.J.~Olson, C.~Reichhardt, and S.~Bhattacharya, Phys.~Rev.~B {\bf 64},
024518 (2001). 

\bibitem{Shobo}
S.~Bhattacharya and M.J.~Higgins, Phys.~Rev.~Lett.~{\bf 70}, 2617 (1993).

\bibitem{Pardo}
F.~Pardo {\it et al.}, Nature (London) {\bf 396}, 348 (1998).

\bibitem{Wigner}
C.~Reichhardt {\it et al.}, Phys.~Rev.~Lett.~{\bf 86}, 4354 (2001).

\bibitem{Fisher}
L.~Balents and M.P.A.~Fisher, Phys.~Rev.~Lett.~{\bf 76}, 2782 (1996).

\bibitem{Lilly}
K.B.~Cooper {\it et al.}, Phys.~Rev.~B, {\bf 60}, R11 285 (1999).

\bibitem{Gronbech}
N.~Gr{\o}nbech-Jensen, Int.~J.~Mod.~Phys.~C {\bf 7}, 873 (1996).

\end{references}
\end{document}